\documentclass[fleqn,usenatbib]{mnras_mod}

\usepackage[T1]{fontenc}
\usepackage{ae,aecompl}

\RequirePackage{rotating}

\usepackage{graphicx}	%
\usepackage{amsmath}	%
\usepackage{amssymb}	%

\usepackage[flushleft]{threeparttable}
\usepackage{float}
\usepackage{datetime}
\usepackage{textpos}
\usepackage{booktabs}
\usepackage{rotating}
\usepackage{adjustbox}
\usepackage{epsfig}
\usepackage{textcomp}
\usepackage{color}
\usepackage{bm}

\usepackage[all]{hypcap}

\usepackage{breqn}

\providecommand{\adsurl}[1]{\href{#1}{ADS}}

\providecommand{\url}[1]{\href{#1}{#1}}
\usepackage{url}

\usepackage{placeins}

\def \aj{Astronom.~J.}

\def \aap{A\&A }
\def \apjl{ApJ}
\def \apjs{ApJS}
\def \apj{ApJ}
\def \jcap{JCAP}

\def \mnras{MNRAS}

\def\alt{\raise0.3ex\hbox{$\;<$\kern-0.75em\raise-1.1ex\hbox{$\sim\;$}}}
\def\agt{\raise0.3ex\hbox{$\;>$\kern-0.75em\raise-1.1ex\hbox{$\sim\;$}}}

\newcommand{\bw}{\begin{widetext}}
	\newcommand{\ew}{\end{widetext}}

\newcommand{\lsim}{\,\rlap{\raise 0.35ex\hbox{$<$}}{\lower 0.7ex\hbox{$\sim$}}\,}
\newcommand{\gsim}{\,\rlap{\raise 0.35ex\hbox{$>$}}{\lower 0.7ex\hbox{$\sim$}}\,}

\def\lesssim{\mathrel{\hbox{\rlap{\hbox{\lower3pt\hbox{$\sim$}}}\hbox{\raise2pt\hbox{$<$}}}}}
\def\gtrsim{\mathrel{\hbox{\rlap{\hbox{\lower3pt\hbox{$\sim$}}}\hbox{\raise2pt\hbox{$>$}}}}}

\def\xlinkspace#1 #2{%
	\ifx\relax#2%
	\xlinkdash#1-\relax
	\else
	\xlinkdash#1 -\relax
	\expandafter\xlinkspace\expandafter#2%
	\fi}

\def\xlinkdash#1-#2{%
	\ifx\relax#2%
	\tmp{#1}%
	\else
	\tmp{#1-}%
	\expandafter\xlinkdash\expandafter#2%
	\fi}

\DeclareMathOperator{\sech}{sech}

\newcommand{\newtext}[1]{\textcolor{black}{#1}}
\newcommand{\newnewtext}[1]{\textcolor{black}{#1}}

\title[Density Reconstruction of the Galactic Bulge Area]{
Non-Parametric Density Reconstruction of the Galactic Bulge Area 
using Red Clump Stars in the VVV Survey}

\author[Paterson et al.]{
		\mbox{ Dylan~Paterson\thanks{E-mail: dnp16@uclive.ac.nz},
		Brendan~Coleman,  Chris~Gordon} 
\\
 School of Physical and Chemical Sciences, University of Canterbury, Christchurch, New Zealand\\
}

\date{}

\pubyear{}

\begin{document}
\label{firstpage}
\pagerange{\pageref{firstpage}--\pageref{lastpage}}
\maketitle

\begin{abstract}
Studies of the red clump giant population in the inner Milky Way suggest the Galactic bulge/bar has a boxy/peanut/X-shaped structure as predicted by its formation via a disc buckling instability. We used a non-parametric method of estimating the Galactic bulge morphology that is based on maximum entropy regularisation. This enabled us to  extract the three-dimensional distribution of the red giant stars in the bulge from deep photometric catalogues of the VISTA Variables in the Via Lactea (VVV) survey. Our high-resolution reconstruction confirms the well-known boxy/peanut/X-shaped structure of the bulge. We also find spiral arm  structures that extend to around three kpc in front of and behind the bulge and are on different sides of the bulge major axis. \newnewtext{We show that the detection of these structures is robust to the uncertainties in the luminosity function. }

\end{abstract}


\section{Introduction}

Extragalactic studies of disc galaxies have found bulges/bars with boxy/peanut/X-shaped (B/P/X) components are relatively common in nearby early type (S0 - Sd) disc galaxies \citep{LaurikainenMilkyWaymass2014,CiamburGraham2016}. 
\newtext{ Theories  of  secular  evolution  suggest  that  bulges  can undergo a thickening out of the plane driven by a vertical resonance, evolved more rapidly by buckling, resulting in this B/P/X shape \cite{Sellwood2014SecularEvolution}.}
These buckling processes have been observed in N-body simulations of disc galaxies, where the resulting bulge exhibits a strong B/P/X geometry \citep{Bureau2005,Debattista2006}. When viewed at an inclined angle, the bars of N-body simulated galaxies in the post buckling phase appear to have offset spurs at the end, which have also been observed in other galaxies  \citep{Erwin2013PeanutsatanAngle,Erwin2016CaughtintheAct}. 
\newtext{The offset of these spurs are associated with the presence of bulge components such as the long bar and B/P/X bar in the Milky Way. }
However, a B/P/X geometry does not automatically mean that the Galactic bar has buckled as the B/P/X bulge may have formed via an orbital resonance \citep{Quillen2014}.

Studying the bulge of the Milky Way is challenging due to our viewing angle of the Galactic bulge. One promising method to uncover the shape of the inner Milky Way is to use the red clump (RC) stars, which have a narrow intrinsic luminosity range, as standard candles \citep{GirardiRedClumpStars2016}.

The RC has been the focus of several studies characterising the three-dimensional density structure of the Galactic bulge. The most common class of parametric model used to describe the bulge is the triaxial ellipsoid \citep{StanekModellingGalacticBar1997,RattenburyModellingGalacticbar2007,Caonewphotometricmodel2013,Simionparametricdescription3D2017}. Although these triaxial models do a reasonable job of describing the general structure of the bulge, some of the studies, e.g \cite{Simionparametricdescription3D2017} (hereafter S17), show evidence of non-triaxial structures in the residual star-counts. Non-parametric methods have also been used involving deconvolution or constant intrinsic RC magnitude assumptions as in \cite{WeggMappingthreedimensionaldensity2013}  and \cite{SaitoMappingXshapedMilky2011} respectively.
The Galactic RC was found to produce a double photometric peak 
by
\cite{NatafSplitRedClump2010} using  OGLE-III data and \cite{McWilliamTwoRedClumps2010} using 2MASS. 
The deconvolution results of \cite{WeggMappingthreedimensionaldensity2013} (from here on WG13) showed a B/P/X bulge using the RC stars in the VVV survey.

We developed and applied our fully non-parametric deprojection of the Galactic bulge stars, relying only on choice of luminosity function and smoothness regularisation. In applying the principle of maximum entropy for statistical inference \citep{JaynesInformationTheoryStatistical1957}, %
we aimed to produce a smooth density estimate of the Galactic bulge region and explore potential features of interest. In particular, we are interested in features such as the X-shape of the bulge and spiral arm structures on the ends of the bulge, which are difficult to model with existing methods. 

Our article is arranged as follows:
In Sections \ref{sec:Data-section} and \ref{subsec:luminosity function} we outline our data selection and we rationalise the choice of data cuts and masks.
Our semi-analytic luminosity function is constructed and its foundations presented in Section \ref{subsec:luminosity function}.
In Section \ref{sec:Method} we present our non-parametric deconvolution method for inverting stellar statistics to recover the three-dimensional stellar density distribution.
We perform maximum entropy deconvolution on the VVV data and interpret the results in Section \ref{sec:spiralrevisited}.

\section{VVV Data}\label{sec:Data-section}

We used data from the MW-BULGE-PSFPHOT compilation \citep{SurotMappingstellarage2019}, an ultra deep, infra-red, photometric catalogue of almost 600 million stars in the Milky Way bulge. Included in the catalogue are $K_s$ and $J$ apparent magnitudes from PSF fitting VVV images \citep{MinnitiVISTAVariablesLactea2010}, completeness for most stars from artificial star tests, extinction corrected $K_s$ and $J$ magnitudes, combined photometric + systematic uncertainties for $K_s$ and $J$, and a variety of quality metrics. 

From this catalogue we constructed binned star counts on a $(80 \times 100 \times 75)$ linear grid in extinction corrected magnitude $(K_s$), Galactic latitude ($l$), and Galactic longitude ($b$). 
The range of the grid was 11 < $K_s$ < 15, $-10^{\circ} < l <  10^{\circ}$, and  $-10^{\circ} < b < 5^{\circ}$.
\newtext{This resulted in a corresponding voxel size of 
$0.05\mbox{ mag}\times12\mbox{ arcmin}\times12\mbox{ arcmin}$.}
To select mainly the red giant stars, we excluded sources with  $0.4 < J - K_s < 1.0$. 
A few sources in the catalogue do not have completeness values, as the detectors on which they were observed were excluded from the completeness analysis, so we were unable to completeness correct our star counts on a star-by-star basis. Instead, we calculated the mean completeness in each $(K_s,l,b)$ voxel. 
 We corrected for completeness by dividing the number count of stars in a voxel by the estimated completeness of that voxel.

\begin{figure*}
    \centering
    \includegraphics[width=\textwidth]{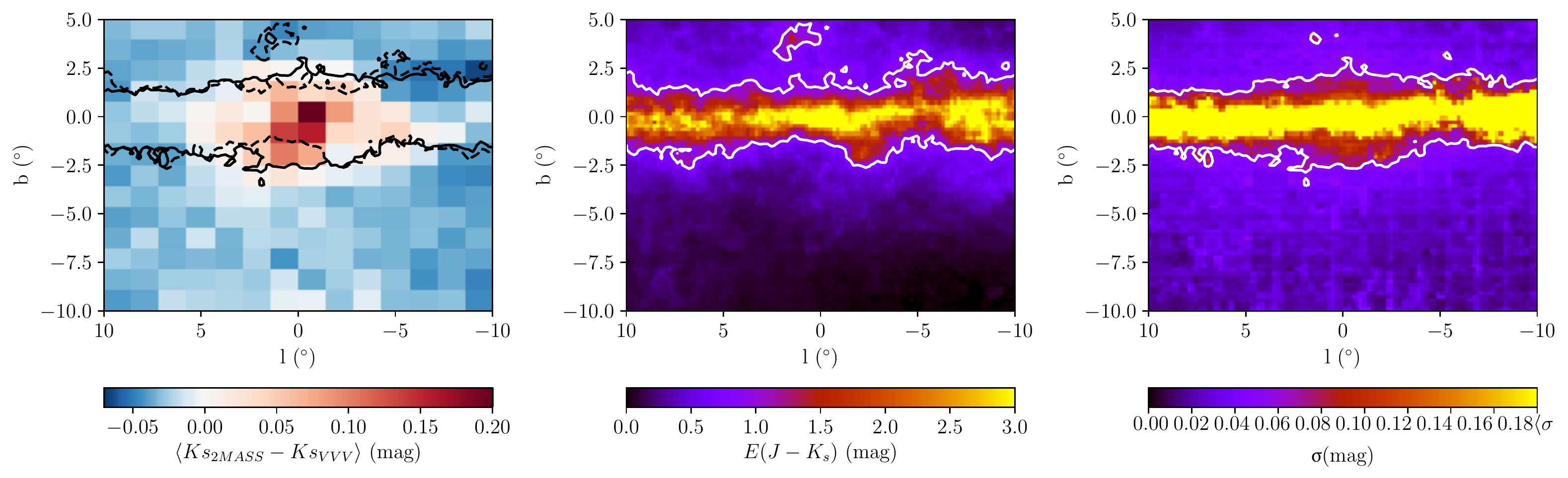}
    \caption{\textit{Left:} Median difference in $K_s$ between cross matched 2MASS and VVV sources.  We used this difference to correct the photometric zero-point within each tile. 
    The solid line is the  
    \newtext{$\sigma=0.06$}
    mask boundary and the dashed line is the  $E(J-K{_s}) = 0.9$ mask boundary. \textit{Middle:} Color excess used in the extinction correcting the MW-BULGE-PSFPHOT photometry. Inside the white boundary, $E(J-K_s) > 0.9$, extinction severely degrades the quality of the VVV photometry. \textit{Right:} Mean over $12.975<K_s<13.025$ of the combined photometric and systematic $K_s$ uncertainty from the PSF fitting procedure used in compiling the MW-BULGE-PSFPHOT catalogue.
    Inside the white boundary,
    \newtext{$\sigma>0.06$},
    the photometry is affected by the increased crowding, causing blending and source confusion.}
    \label{fig:vvvmasks}
\end{figure*}

The photometry in the MW-BULGE-PSFPHOT compilation was calibrated relative to the Cambridge Astronomical Survey Unit (CASU) aperture photometry catalogues \citep{SaitoVVVDR1}, which are known to have field-to-field variations in $K_s$ zero-point of up to 0.1 mag
 \newtext{\citep{Hajdu2019OptimalVVVCalibration}}. 
We corrected for this variation in zero-point by adding to the Ks magnitudes, within each tile, the median difference between the 2MASS point source catalogue \citep{SkrutskieTwoMicronAll2006} Ks magnitude and the non extinction corrected MW-BULGE-PSFPHOT Ks magnitude.
We limited the cross matching to sources in 2MASS with 12 < $K_s$ < 13 to ensure good photometric quality in both source catalogues and used a cross matching threshold of $0.1^{\prime\prime}$. This limit was used  to reduce to effect of crowding and source merging in the 2MASS catalogue \citep{Hajdu2019OptimalVVVCalibration}. The photometric offsets are shown on the left panel of Fig. \ref{fig:vvvmasks}. 

Even after extinction correction and completeness correction, some regions on the sky had residual effects in their star counts. We chose to exclude these regions from our analysis by masking where the crowding and extinction is high
\newtext{To do this we evaluated the  total error in the $K_s$ measurement in the same way as done by WG13: 
\begin{equation}
    \sigma=\sqrt{\left<\sigma_{K_s}\right>^2+\left<A_{K_s}\sigma_{E,JK}\right>^2}
    \label{eq:totalerror}
\end{equation}
where the photometric error ($\sigma_{K_s}$), the residual reddening ($\sigma_{E,JK}$), and the extinction ($A_{K_s}$) is provided or can be inferred for each star in the   MW-BULGE-PSFPHOT  compilation. The angular brackets in Eq.~\ref{eq:totalerror} denote the average value over a voxel.}
In the right panel of Fig. \ref{fig:vvvmasks}, the mean $K_s$ magnitude error 
\newtext{$\sigma$}
of stars with $12.975 < K_s < 13.025$ is shown. The value of the exclusion boundary, 
\newtext{$\sigma = 0.06$}, was chosen to visually match the $E(J-K)=0.9$ boundary in the less crowded $|l|>5^{\circ}$ region. We can see that in the left panel of Fig. \ref{fig:vvvmasks},
our 
\newtext{$\sigma$}
based mask
mask  excluded from the analysis nearly all the tiles with a significant positive photometric correction. Pixels that contain globular clusters in the GLOBCLUST \citep{Harris2010GlobularClusterCatalogue} globular cluster catalogue were also excluded from the analysis.

\section{Isochrones, Bulge Metallicity and Luminosity functions}\label{subsec:luminosity function}

Previous studies (e.g.~WG13 and S17)
have produced luminosity functions by fitting a parametric model to simulated populations of stars, with masses randomly drawn from initial mass functions (IMFs) and absolute magnitudes assigned by interpolation of mass-absolute magnitude relations from isochrones. In this framework, the absolute magnitude and mass are treated as random variables, where the luminosity function is the probability density function of absolute magnitude, and the IMF is the probability density function of mass. Instead of simulating the luminosity function, we adopted a more analytic approach. The luminosity function for a specific age $\tau$ and metallicity $z$ is determined by

\begin{equation}
    \phi \left ( M_{K_{s}}, z, \tau \right ) = \sum_i{\xi\left (\theta_i^{-1}\left ( M_{K_{s}},z,\tau\right ) \right)\left| \frac{d\theta_i^{-1}\left ( M_{K_{s}},z,\tau \right )}{d M_{K_{s}}} \right|}
    \label{eq:luminosity functionimfrelation}
\end{equation}
where $\xi$ is the IMF and $\theta$ is the mass-absolute magnitude relation 
\begin{equation}
    M_{K_{s}} = \theta\left(m,z,\tau\right).
    \label{eq:massabsrelation}
\end{equation}
In mass ranges where $\theta$ is not uniquely invertible, the luminosity function is summed over all possible solutions to the inversion of $\theta$. To get the luminosity function for the full population, we took the expected value of Eq.~\eqref{eq:luminosity functionimfrelation}  
\begin{equation}
    \Phi \left ( M_{K_{s}}\right ) = \int^{\infty}_{-\infty}\int^{\infty}_{-\infty} \phi\left( M_{K_{s}},z,\tau\right) f\left(z,\tau\right) dz d\tau.
    \label{eq:luminosityfunction}
\end{equation}
where $f$ is the metallicity distribution function. We assumed a bulge age of 10 Gyr with metallicity normally distributed with solar mean metallicity $\mu_{\text{\lbrack Fe/H \rbrack}} = 0.0 $ and metallicity dispersion $\sigma_{\left \lbrack \text{Fe/H} \right \rbrack} = 0.4$ \citep{Zoccali2008BulgeMetalContent}. 

\begin{figure}
    \centering
    \includegraphics[width=0.95\columnwidth]{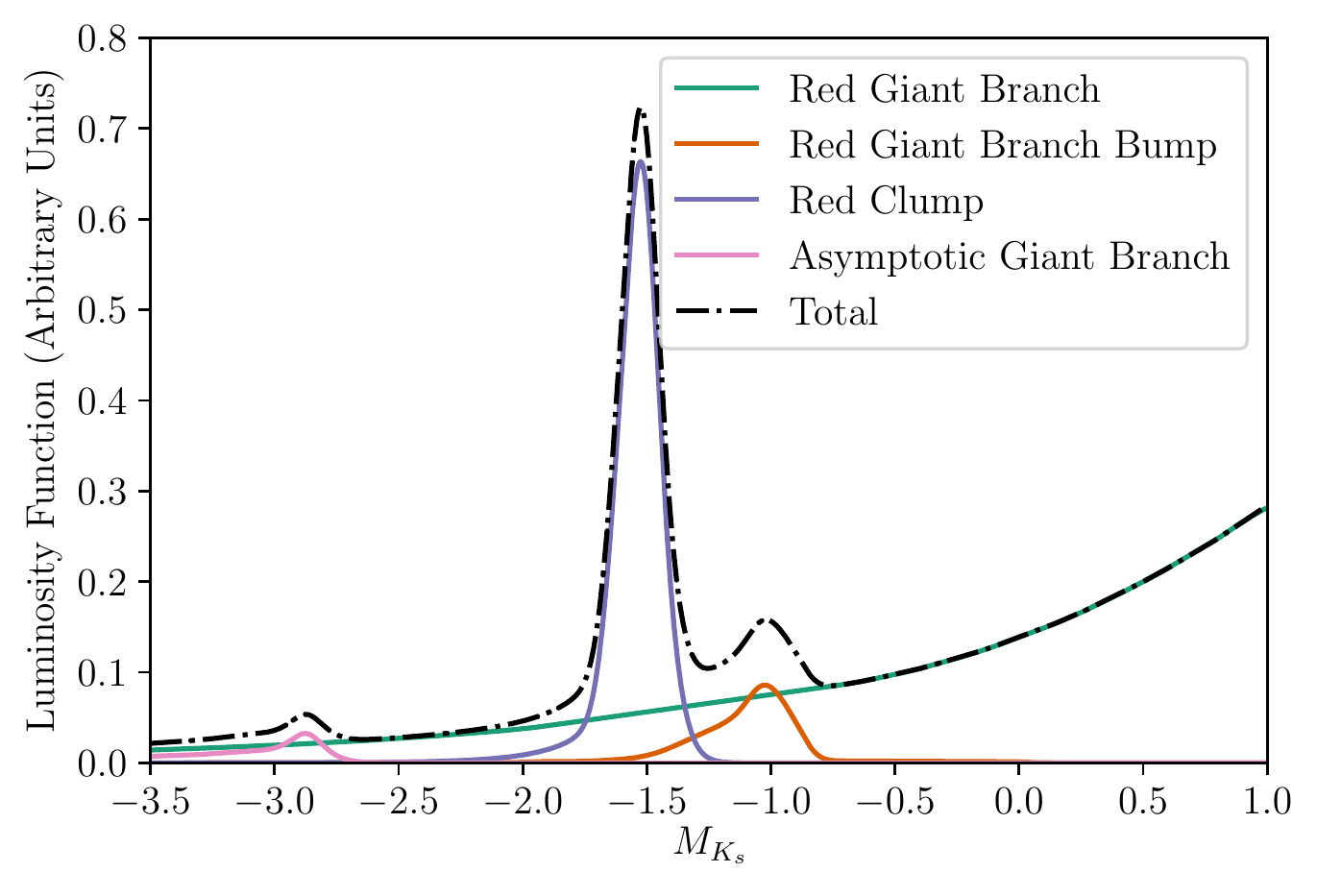}
    \caption{Luminosity function of a 10 Gyr old population
       with $\langle \left \lbrack \text{Fe/H} \right \rbrack \rangle = 0.0 $ and $\sigma_{\left \lbrack \text{Fe/H} \right \rbrack} = 0.4$ using PARSEC+COLIBRI isochrones and \protect\cite{ChabrierGalacticStellarSubstellar2003} 
     log-normal IMF. We convolved the luminosity function by a Gaussian with standard deviation equal to the combined photometric and systematic uncertainty in $K_s$. For display purposes, the luminosity function, in this figure, was convolved with a Gaussian with $\sigma=0.05$ which is a typical value for the error in $K_s$.}
    \label{fig:luminosityfunction}
\end{figure}

We constructed our bulge luminosity function using mass-absolute magnitude relations from the PARSEC+COLIBRI 10 Gyr isochrone sets \cite{MarigonewgenerationPARSECCOLIBRI2017} using 39 metallicity bins linearly spaced in the range -2.279 < [Fe/H] < 0.198. These isochrones are tabulated at fixed mass and metallicity values. The magnitude values between the fixed points were interpolated using a linear univariate spline in mass along a single metallicity isochrone. Attempting to interpolate between evolutionary stages where there are large changes in luminosity, e.g.\  first ascent red giant to helium core burning giant, introduced artefacts in the resulting luminosity function, so we used the evolutionary stage flags in the isochrones to separate them; 0-3 red giant branch, 4-6 RC and $>6$ asymptotic giant branch.   Fig.~\ref{fig:luminosityfunction} shows the luminosity function calculated using Eq.~\eqref{eq:luminosityfunction} with mass-absolute magnitude relations from PARSEC+COLIBRI isochrones and a 
\cite{ChabrierGalacticStellarSubstellar2003} 
log-normal IMF. Fitting a Gaussian to the RC component gave a mean absolute magnitude $\mu_{M_{K_s}} = -1.53$ with standard deviation $\sigma_{RC} = 0.06$ which is consistent with the luminosity function of S17. 

Observational effects such as residual extinction and crowding introduce uncertainty in measuring the $K_s$ apparent magnitude, which effectively broadens the observed luminosity function. 
\newtext{ We accounted for this by convolving our semi-analytic luminosity, described above, with a zero-mean Gaussian
which had a standard deviation, $\sigma$, which is given in Eq.~\ref{eq:totalerror}.  }

\section{Deconvolution Method}\label{sec:Method}

The stellar density ($\rho$) of the Galactic bulge can be reconstructed by inverting the equation of stellar statistics (e.g.~\cite{Lopez-CorredoiraInversionstellarstatistics2000}, WG13)
\begin{multline}\label{eq:stellarstatistics}
    N\left(K_{s},l,b\right) =N_{\rm{thin}}\left(K_{s},l,b\right) + N_{\rm{thick}}\left(K_{s},l,b\right) \\+ \Delta \Omega \Delta K_s \int^{13}_{4} \rho\left(s,l,b\right) \Phi\left(K_{s} - 5\log s - 10\right) s^2 \, {\rm d}s
\end{multline}
where $N$ denotes total number of stars
in a voxel centred at $(K_s,l, b)$.
This is made up of contributions from the the thin disc ($N_{\rm{thin}}$), thick disc ($N_{\rm{thick}}$),
and a contribution given by a
weighted integral of the Galactic bulge number density ($\rho$).
The
 $\Delta \Omega$ denotes the solid angle subtended by the line-of-sight, $\Delta K_s$ denotes the width of a $K_s$ magnitude bin, and and  $s$ denotes the distance from the Sun measured in kpc.
 An example of the the broadened luminosity function ($\Phi$) is shown in Fig.~\ref{fig:luminosityfunction}.
We chose the  integration
range $4~{\rm kpc}\leq s\leq 13~{\rm kpc}$ as the Galactic bulge density is negligible outside that region.

Following S17, we modelled the thick and thin discs using the description for the Besan\c{c}on galaxy model \citep{Robinsyntheticviewstructure2003}. The thin disc was constructed from seven sub-populations which have different ages spanning 0-10 Gyr, where the star formation rate was assumed constant for each sub-population. All sub-populations were assumed to have relaxed into isothermal distributions, where the density distribution is a cylindrically symmetric holed ellipsoid, described by an  \cite{Einasto1979galacticmassmodelling} density law 
\begin{equation}\label{eq:thindiskdensity}
    \begin{aligned}
            \rho_{\rm thin}(R,Z_{\rm cyl}) = \rho_0  \left\lbrack \exp \left( -\sqrt{0.25+ \left(\frac{a}{h_t}\right)^2 }\right) \right. \\
            \left. - \exp \left( -\sqrt{0.25+ \left(\frac{a}{h_h} \right)^2} \right) \right\rbrack
    \end{aligned}
\end{equation}
with
\begin{equation}
    a^2 = R^2 +  \left(\frac{Z_{\rm cyl}}{\epsilon}\right)^2
\end{equation}
where $R$ and $Z_{\rm cyl}$ are cylindrical co-ordinates in kpc, $h_t$ is the scale length of the disc and $h_h$ is the scale length of the hole in kpc. The axis ratio of the ellipsoid, $\epsilon$, is age dependent. The values for the thin disc density parameters from \cite{Robin2012thindiskupdate} were used in this model, and are summarised in Table. \ref{tab:diskdensity}. We generated a luminosity function for each sub-population of the thin disc using the method described in Section \ref{subsec:luminosity function} assuming a broken power law IMF
\begin{equation}\label{eq:IMFthindisk}
    \xi \left( m \right ) = \left\{\begin{array}{lr}
    m^{-1.6}, & m \leq 1 M_\odot \\
    m^{-3.0}, & m > 1 M_\odot\\
    \end{array}\right.
\end{equation}
Within each sub-population, the metallicity is distributed normally in $\lbrack \rm{Fe/H}\rbrack$ with mean and dispersion as given in Table \ref{tab:diskmetallicity}. We used mass-absolute magnitude relations from the PARSEC+COLIBRI isochrones \citep{MarigonewgenerationPARSECCOLIBRI2017}.

The formation history of the thick disc was assumed to be a single burst event 12 Gyrs ago. The density profile used is distributed exponentially radially, where vertically the density is parabolic near the plane, transitioning to exponential further away from the plane
\begin{multline}
\label{eq:thickdiskdensity}
    \rho_{\rm thick}\left(R,Z \right) = \\ \left\{\begin{array}{lr}
    \rho_0 \exp \left ( -\frac{R - R_\odot}{h_T} \right) \left\lbrack 1 - \frac{Z^2}{h_z}\frac{1}{\zeta\left(2 + \zeta/h_Z \right)} \right \rbrack & Z \leq \zeta \\
    \rho_0 \exp \left ( -\frac{R - R_\odot}{h_T} - \frac{|Z - Z_\odot|}{h_Z} \right) \frac{2 \exp \left( \zeta/h_Z \right) }{1 + \zeta/h_Z} & 
    Z > \zeta \\
    \end{array}\right.
\end{multline}
where $(R_\odot,Z_\odot)=(8.0 \text{ kpc}, 15 \text{ pc})$ is the position of the Sun. Parameter, $h_T$ is the radial scale length, $h_Z$ is the vertical scale height and $\zeta$ is the height where the density transitions from parabolic to exponential. The IMF for the thick disc is a simple power law
\begin{equation}\label{eq:IMFthickdisk}
    \xi \left( m \right ) =  m^{-0.22}.
\end{equation}
Both the thick and thin discs were modelled as having a warp and a flare,
\begin{equation}\label{eq:warp}
    Z_{\text{warp}} = \gamma_{\text{warp}}(R - R_{\text{warp}}) \cos \left( \phi - \phi_{\text{warp}} \right)
\end{equation}
where the density in Eq.~\eqref{eq:thindiskdensity} and Eq.~\eqref{eq:thickdiskdensity} at $Z$, is instead evaluated at $Z + Z_{\text{warp}}$ when $R > R_{\text{warp}}$; $\phi_{\text{warp}}$ is the direction in which the warp is maximum. The flare was modelled by linearly increasing the scale height by
\begin{equation}\label{eq:flare}
    h_{\text{flare}} = \gamma_{\text{flare}}(R - R_{\text{flare}})
\end{equation}
when $R > R_{\text{flare}}$. We used the same parameters for the flare and warp as \cite{Robinsyntheticviewstructure2003}; $\gamma_{\text{warp}}= 0.18$, $R_{\text{warp}} = 0.98 R_{\odot}$, $\phi_{\text{warp}} = 90.0^{\circ}$, $\gamma_{\text{flare}}= 0.0054$ and $R_{\text{flare}} = 1.12 R_{\odot}$. The disc parameters we used are listed in Tables~\ref{tab:diskdensity} and \ref{tab:diskmetallicity}.

\begin{table}
\label{tab:diskdensity}
    \begin{center}
    \caption{Density distribution parameters for the Besan\c{c}on thick and thin discs }

    \begin{tabular}{ |c|c|c|c|c| }
    
        \hline
        Component & Age &$h_{t/T}$ & $\epsilon/h_Z$ & $h_h$ \\ 
        \hline
        Thin Disc & 0.0-0.15  & 5.00 & 0.0140 & 3.00  \\
                  & 0.15-1 & 2.53 & 0.0268 & 1.32 \\ 
                  & 1-2 & 2.53 & 0.0375 & 1.32  \\
                  & 2-3 & 2.53 & 0.0551 & 1.32  \\
                  & 3-5 & 2.53 & 0.0696 & 1.32  \\
                  & 5-7 & 2.53 & 0.0785 & 1.32  \\
                  & 7-10 & 2.53 & 0.0791 & 1.32  \\
        \hline
        Thick Disc &  12  & 2.36 & 0.535 & - \\
        \hline
    \end{tabular}
    
    \end{center}
\end{table}
\begin{table}\label{tab:diskmetallicity}
    \begin{center}
    \caption{
    Metallicity distribution parameters for the Besan\c{c}on thick and thin discs }

    \begin{tabular}{ |c|c|c|c| }
    
        \hline
        Component & Age (Gyr) & $\mu_{\lbrack \text{Fe/H}\rbrack}$ & $\sigma_{\lbrack \text{Fe/H}\rbrack}$ \\ 
        \hline
        Thin Disc & 0.0-0.15  & -0.01 & 0.12 \\
                  & 0.15-1 & -0.03 & 0.12 \\ 
                  & 1-2 & -0.03 & 0.10 \\
                  & 2-3 & -0.01 & 0.11 \\
                  & 3-5 & -0.07 & 0.18 \\
                  & 5-7 & -0.14 & 0.17 \\
                  & 7-10 & -0.37 & 0.20 \\
        \hline
        Thick Disc &  12  & -0.78 & 0.3 \\
        \hline
    \end{tabular}

    \end{center}
\end{table}

\subsection{Maximum Entropy Deconvolution}\label{subsec:maxentdeconv}
Maximum entropy methods (MEMs) 
have been used in applications such as image reconstruction in radio interferometry \citep{Cornwell1985MEMRadioInterferometry}.
They have also been used in foreground/background modelling of diffuse emission processes, e.g.\   cosmic microwave background studies with WMAP \citep{2003BennetWMAPYear1Foreground} and diffuse gamma-ray studies with Fermi-LAT \citep{StormSkyFACTHighdimensionalmodeling2017a}. 

We used penalised likelihoods with penalties which come in two general forms: the first is maximum entropy regularisation which is defined for a field $\kappa$, \begin{equation}\label{eq:maxentdef}
    -2 \ln \mathcal{L}_{\rm MEM} = 2 \lambda \sum_{i,j,k} \left( 1 - \kappa_{i,j,k} + \kappa_{i,j,k} \ln \kappa_{i,j,k} \right)
\end{equation}
where $i$, $j$, and $k$ are the grid points for $s$, $l$, and $b$ respectively. 
As shown in the Appendix, $\ln \mathcal{L}_{\rm MEM}$ has an extremum at $\kappa_{i,j,k} = 1$. 
We used a parameterisation where $\kappa$ is the ratio between a modelled quantity of interest and a smooth prior estimation of the quantity. Where there is little information in the data about the modelled quantity, such as noisy or low count regions, the model will tend towards the prior. As shown in the Appendix, the prior uncertainty or dispersion of $\kappa$ is $\lambda^{-1/2}$. So for example, if we expected deviations of around 10\% from the smooth prior estimation of the quantity, we would  set $\lambda=100$.
The larger the value of $\lambda$ chosen, the smaller the prior uncertainty assumed and so the more regularisation of the solution is applied.

The second form of likelihood penalty is $\ell_2$-norm regularisation of the second derivative of the logarithm. If $\rho$ varied over one dimension, we would use the usual second order central difference equation approximation of curvature:
\begin{equation}\label{eq:l2normdef}
  -2 \ln \mathcal{L}_{\rm smooth} = 
\eta \sum_{i}  \left( \ln \rho_{i-1} + \ln \rho_{i+1} - 2\ln \rho_i  \right)^2.
\end{equation}
This penalty  enforces smoothness as it has a minimum when $\ln \rho$ has zero curvature which
occurs for $\ln \rho$ which is constant or varies linearly with $i$. In which case $\rho$ will either be constant or vary as the exponential of a linear function.
Therefore, where there is no data to influence the fit, such as masked regions, this regularisation will tend to give exponential behaviour. 
As shown in Appendix, the prior relative standard deviation from an exponential of a linear function is approximately $1/\sqrt{6\eta}$.
So, the larger the value chosen for $\eta$ the more smoothness regularisation is applied.
A similar smoothness regularizing term was used by \cite{BissantzSpiralarmsbar2002} to estimate the morphology of the bulge from the {\em COBE 
DIRBE\/} data.

Our maximum entropy method constructs a model for predicting the binned star counts, using a non-parametric description of the density. It maximises the  penalised log likelihood:
\begin{equation}\label{eq:likelihoodmaxent}
   \begin{aligned}
         \ln \mathcal{L} = & \sum_{\{i,j,k\}\in \{K_s,l,b\}} \left( n_{i,j,k} \ln N_{i,j,k} - N_{i,j,k}\right)
         \\ 
         &
          -\sum_{\{i,j,k\}\in \{s,l,b\}} \left[\vphantom{\left( \ln \rho_{i,j,k-1} + \ln \rho_{i,j,k+1} - 2\ln \rho_{i,j,k}  \right)^2}
         \lambda \left( 1 - \kappa_{i,j,k} +\kappa_{i,j,k} \ln \kappa_{i,j,k} \right)\right. \\ 
         &\left. +\eta_{s}  \left( \ln \rho_{i-1,j,k} + \ln \rho_{i+1,j,k} - 2\ln \rho_{i,j,k}  \right)^2/2\right.
         \\ 
         &\left. +\eta_{l}  \left( \ln \rho_{i,j-1,k} + \ln \rho_{i,j+1,k} - 2\ln \rho_{i,j,k}  \right)^2/2
         \right.
         \\ 
         &\left. +\eta_{b}  \left( \ln \rho_{i,j,k-1} + \ln \rho_{i,j,k+1} - 2\ln \rho_{i,j,k}  \right)^2/2
         \right ],
   \end{aligned}
\end{equation}
where the first term on the RHS is the log of the Poisson likelihood with the observed counts, $n$, and the predicted counts $N$. The second line has the maximum entropy regularisation term
of the form given in Eq.~\eqref{eq:maxentdef}, where $\kappa$ is the ratio between the fitted stellar density, $\rho$, and a prior estimate of the density, $\rho_{\rm prior}$:
\begin{equation}\label{eq:entropydef}
    \kappa \equiv \frac{\rho}{\rho_{\rm prior}}.
\end{equation}
The last three lines of Eq.~\eqref{eq:likelihoodmaxent} are the smoothness regularisation for the density field, of the form given in Eq.~\eqref{eq:l2normdef},
in the $s$, $l$, and $b$ directions.
Including the maximum entropy term in the likelihood discourages the modelled density from over-fitting to regions of the data that are dominated by noise, where it will instead favour the smooth prior density.
Addition of the smoothness terms discourages spurious high frequency variations in the modelled density by minimising curvature in the logarithm of the density. The smoothness term also has the added benefit of inpainting the density in lines of sight which have been masked out. We set $\lambda=0$ in masked regions so as they are only affected by the smoothness term and the values of the model at the edge of the mask.

 For a smooth prior density of the bulge, we used a parametric S-model \citep{FreudenreichCOBEModelGalactic1998,Simionparametricdescription3D2017}
\begin{equation}
    \rho_{\rm prior} = \rho_0 \sech^2 \left(r_s \right)
    \label{eq:smodel}
\end{equation}
where,
\begin{equation}
    r_s = \left( \left\lbrack \left( \frac{|X|}{x_0} \right)^{c_{\perp}} + \left( \frac{|Y|}{y_0} \right)^{c_{\perp}} \right\rbrack^{\frac{c_{\parallel}}{c_{\perp}}}  + \left(\frac{|Z|}{z_0} \right)^{c_{\parallel}}\right)^{\frac{1}{c_{\parallel}}}
\end{equation}
and $X$, $Y$, and $Z$ are distances measured along a coordinate system that is centered in the bulge and aligned with the bulge axes.

\cite{PaperII} performed extensive tests of our maximum entropy method on a simulated Milky Way population (see our Appendix~\ref{Appendix:sim}). From those tests, we found that a suitable choice of regularisation parameters to reconstruct the stellar bulge density is $\lambda = 0.01$, $\eta_s = 400.0$, $\eta_l = 200.0$, and $\eta_b = 100.0$. We found that the results were insensitive to small changes in these values.

\section{Deconvolution Results}\label{sec:spiralrevisited}

We applied the maximum entropy deconvolution process to the VVV data by maximizing the $\ln\mathcal{L}$ in Eq.~\eqref{eq:likelihoodmaxent}. We used  the Python implementation \textsc{pylbfgs}\footnote{\url{https://github.com/dedupeio/pylbfgs}} of the Limited Memory Broyden-Fletcher-Goldfarb-Shanno (L-BFGS) algorithm.
 The density was modelled non-parametrically on a (257, 100, 75) grid of $(s, l, b)$, in the range $4 < (s/{\rm kpc}) < 13$, $-10^\circ < b < 5^\circ$ and $-10^\circ < l < 10^\circ$, for a total of 1.9275$\times 10^{6}$ free parameters (the grid spacing is $\Delta s, \Delta l, \Delta b$ = 35 pc, 0.2$^{\circ}$, 0.2$^{\circ}$). The normalisation of the thin and thick discs were also left as free parameters.
 To make the optimization of so many parameters feasible, we evaluated the gradients of $\ln\mathcal{L}$ in Eq.~\eqref{eq:likelihoodmaxent} analytically (see Appendix~\ref{Appendix:grad}).
 We used a parametric S model (Eq.~\eqref{eq:smodel}) that was fitted to the VVV data by \cite{PaperII} as the prior density $\rho_{\rm prior}$.

\begin{figure}
\centering
    \includegraphics[width=\columnwidth]{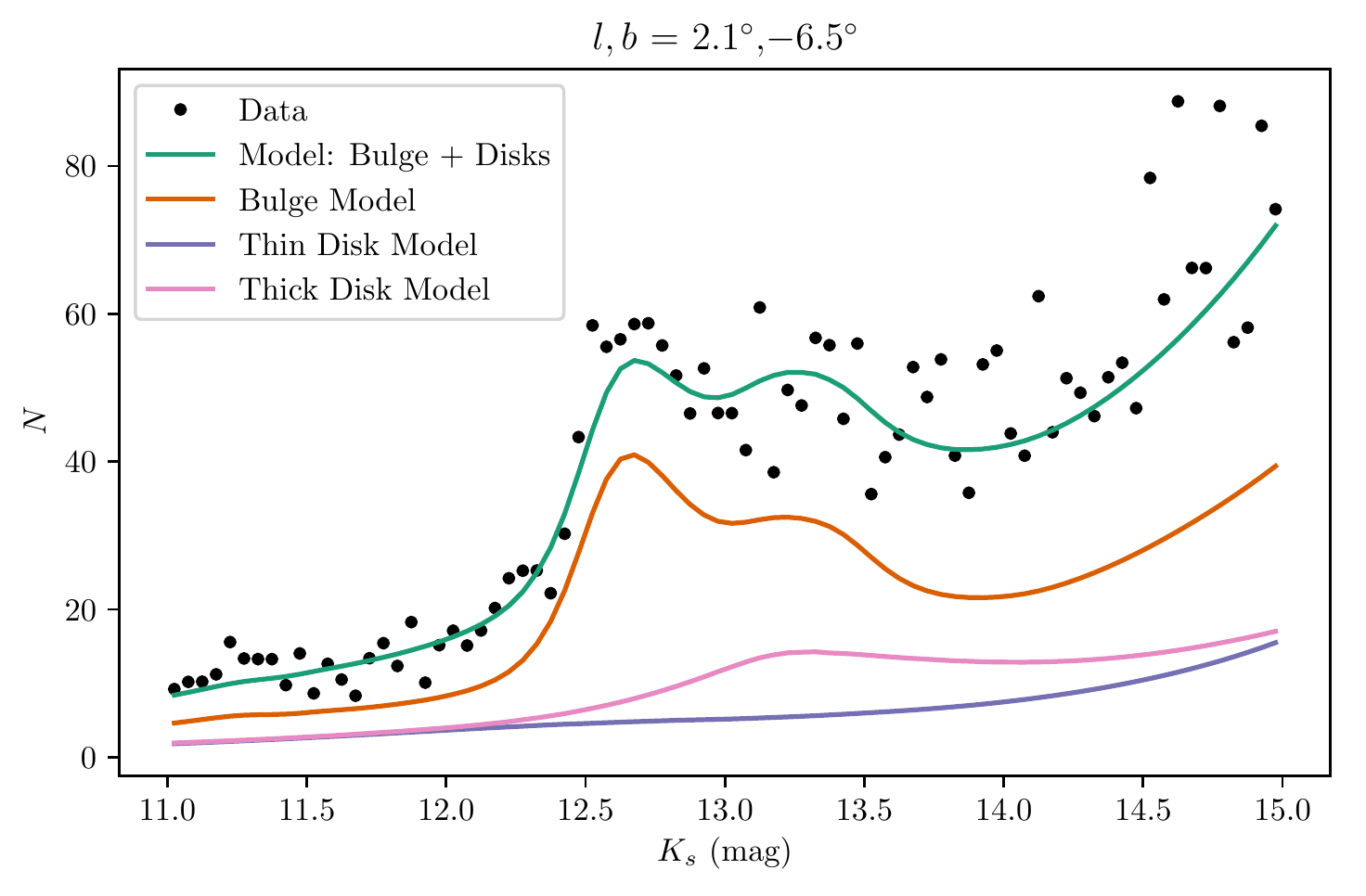}
    \includegraphics[width=\columnwidth]{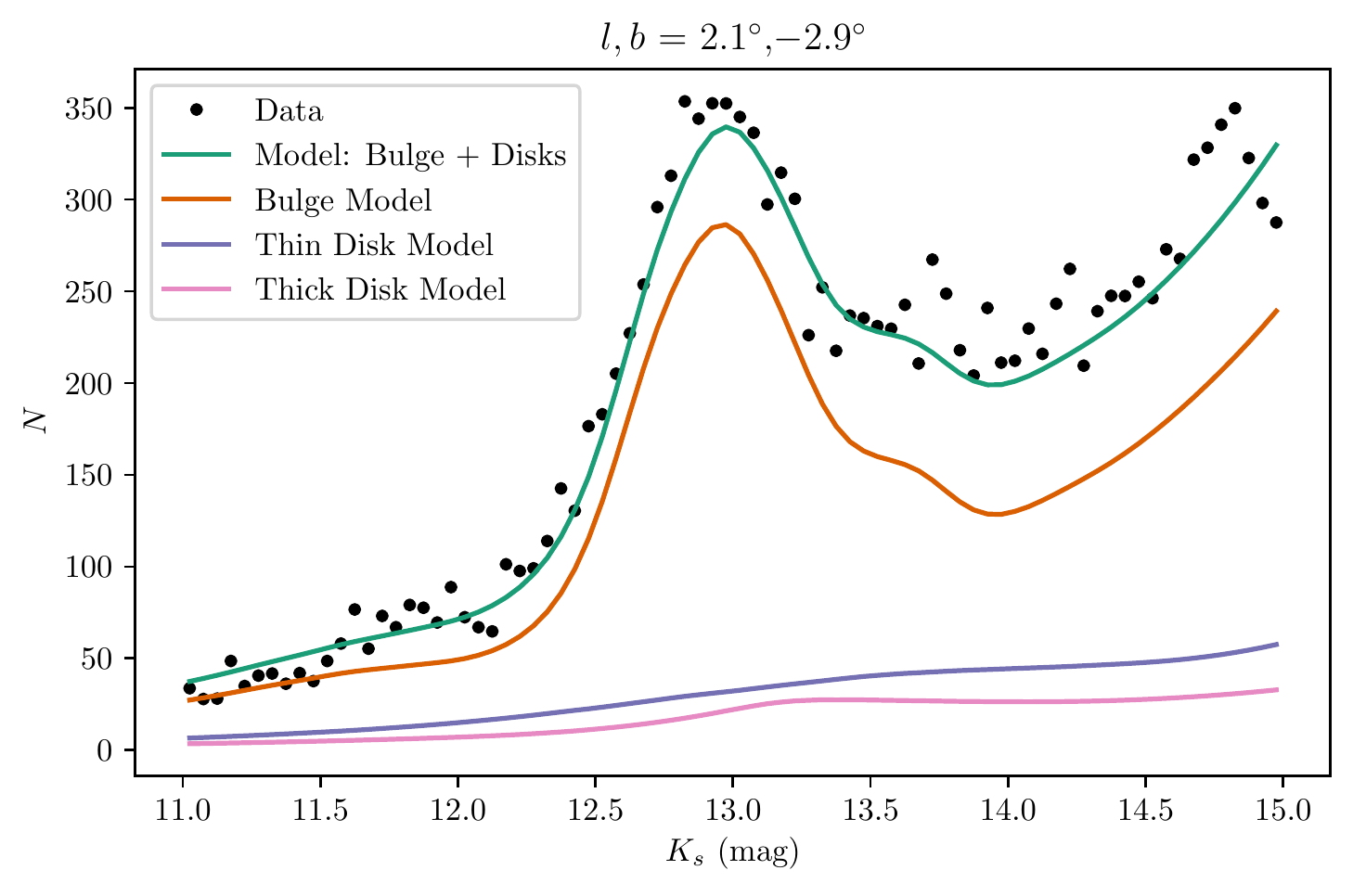}

    \caption{ Demonstration of the fitted model for  a line of sight which displays a splitting in the RC (Top panel) and a line of sight which is near the edge of the masked midplane region (Bottom panel).
    \label{fig:losplots}
    }
\end{figure}

\begin{figure}
\centering
    \includegraphics[width=\columnwidth]{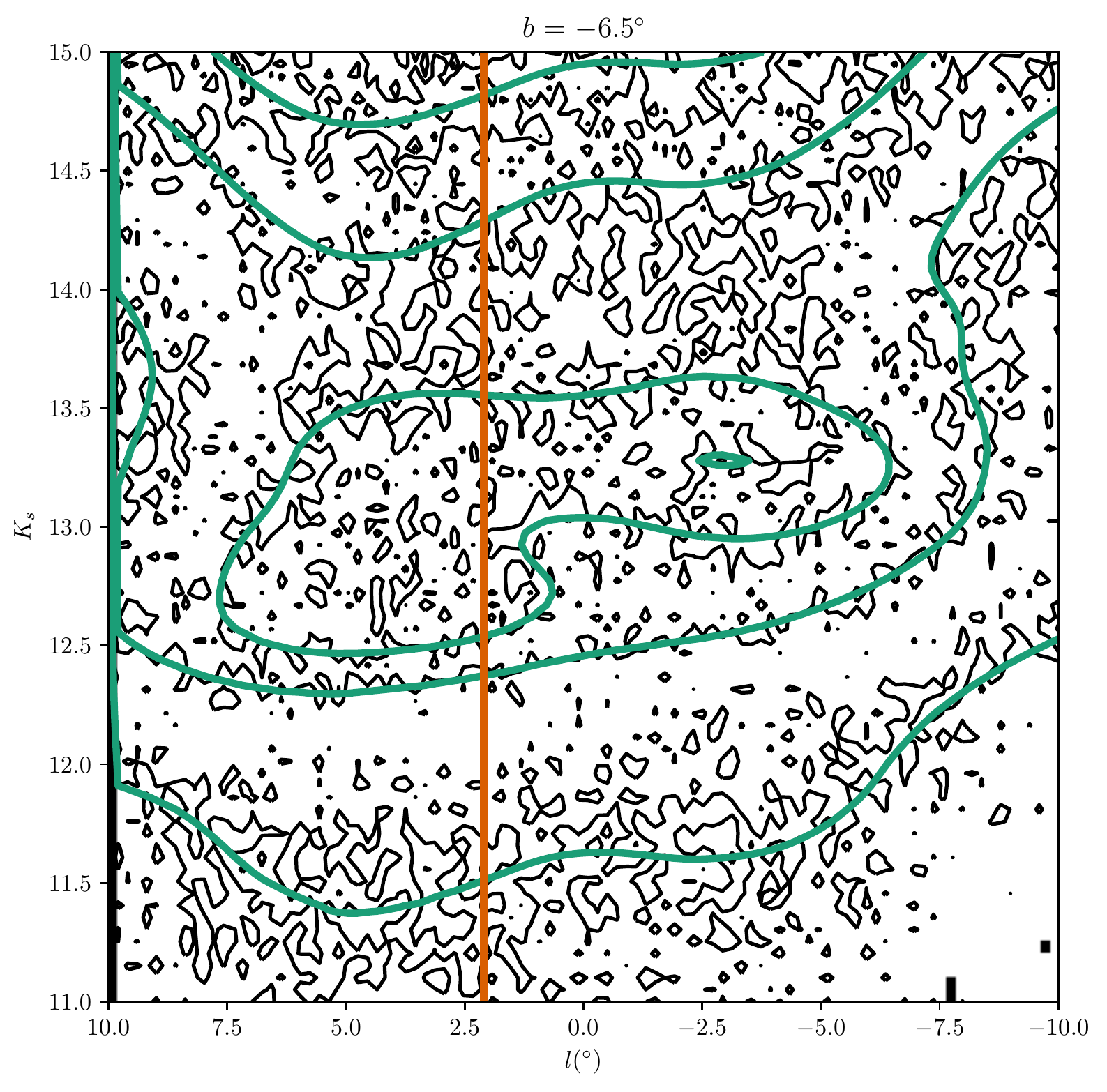}
    \includegraphics[width=\columnwidth]{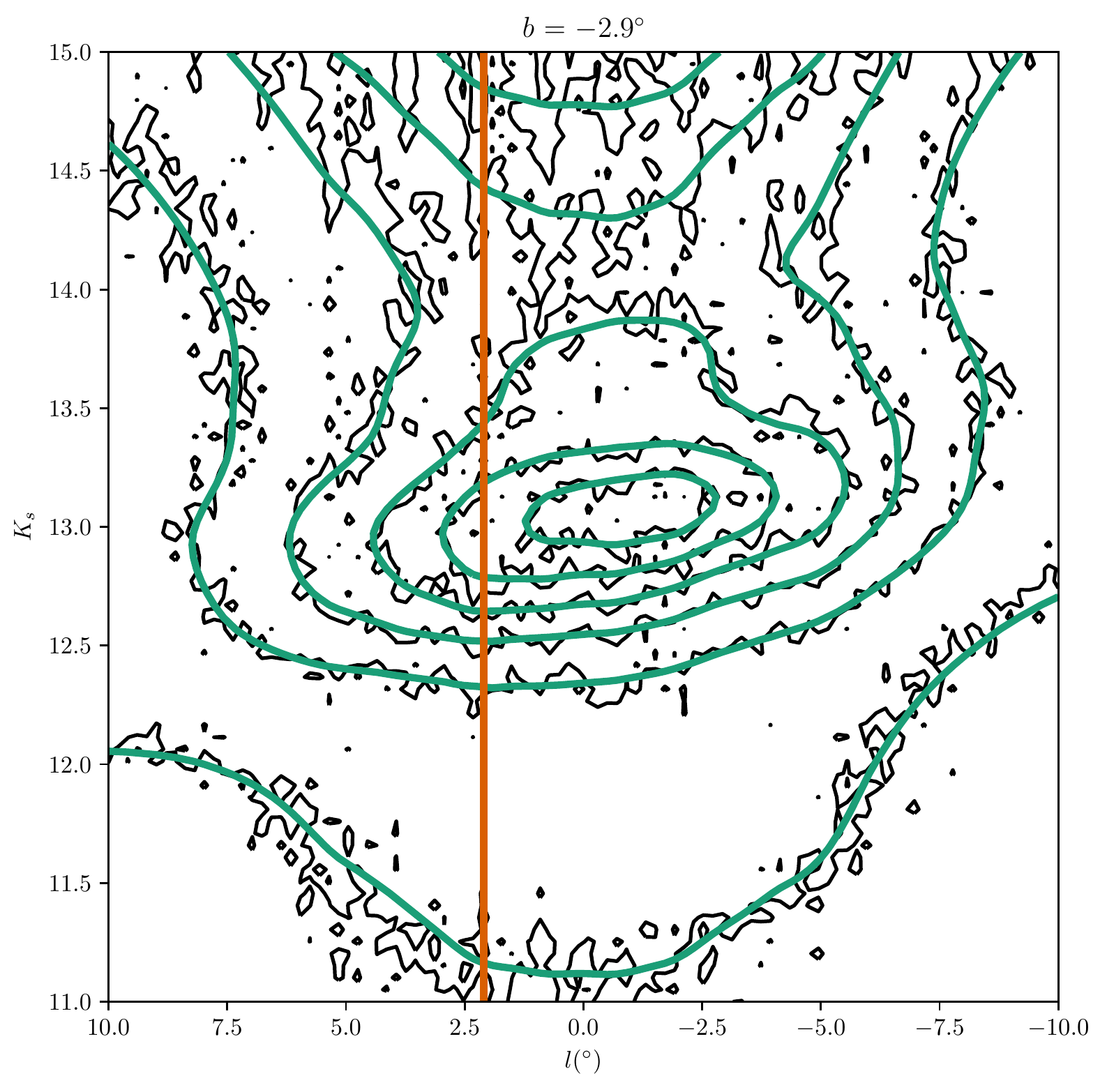}

    \caption{Fitted model (green contours) as compared to the VVV data (black contours) for two representative latitude slices, one that shows the split RC at $b=-6.7^{\circ}$ (top panel), and one that is near the galactic midplane mask at $b=-2.9^{\circ}$ (bottom panel). The line of sights in Fig.~\ref{fig:losplots} are shown in orange.
    \label{fig:contourplots}
    }
\end{figure}

Fig.~\ref{fig:losplots} shows examples of our model fit for two different lines of sight.
At high latitudes, as in the top panel of Fig.~\ref{fig:losplots}, the VVV data is very noisy
due to low number counts.
The maximum entropy method is able to predict the splitting of the RC, even though it is not immediately apparent in the data. Due to the smoothing regularisation, the fitted model for the displayed line of sight is influenced by data in all of the neighbouring voxels. As a result, the model can appear poorly constrained by the data in a single line of sight, as in the bottom panel of Fig.~\ref{fig:losplots}.
When viewed as a slice of the data, as in Fig.~\ref{fig:contourplots}, the model is well constrained across multiple neighbouring line of sights. Due to the narrowness of the RC in the luminosity function, the morphology of the bulge is mainly constrained by the stars in the magnitude range $12 < K_s < 14$. Therefore, the slight bias of the fit at $K_s>14$  is not of particular concern as it is not directly influencing our inferences about the morphology of the bulge region.

\begin{figure*}
\centering
    \includegraphics[width=0.7\textwidth]{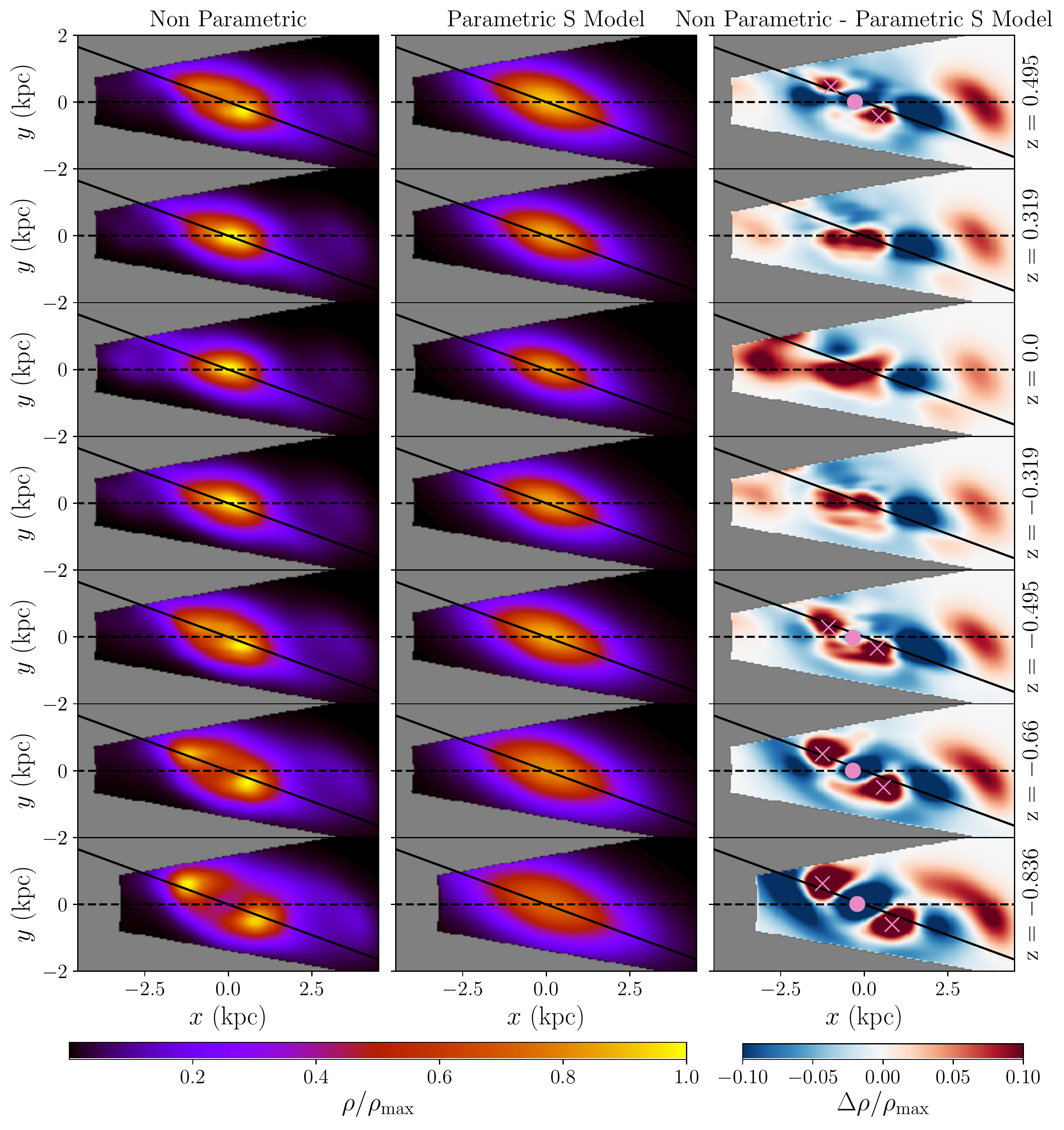}
    \caption{Cartesian projections of the bulge density from the maximum entropy deconvolution (left column) and the parametric prior density of model (middle column). The Sun is located at $(x,y,z)=(-8.0,0.0,0.0)$.  The dashed black line indicates $l=0^{\circ}$ and the solid black line is the major axis of the bulge in the parametric model which is at an angle of $19.8^{\circ}$ from the $l=0^{\circ}$ line. 
    The $z$ coordinate is measured in kpc.
    At $x\sim\pm3$~kpc the spiral arm structures at both ends of the bulge are visible, most clearly in the residuals (right column), which has had the colourbar clipped at $\pm10\%$. The pink crosses indicate the maximum density of the X-arms, and the pink circle is the midpoint between the two arms.
    \label{fig:spiralarmdensity}
    }
\end{figure*}

We plot a Cartesian projection of the reconstructed  bulge density in Fig.~\ref{fig:spiralarmdensity}, where the origin is centred on the maximum density of the bulge. \textcolor{black}{ From this we infer that the Sun is at 
$(x,y,z)=(-8.0,0.0,0.0)$~kpc}.

\newtext{The solid black line in Fig.~\ref{fig:spiralarmdensity} is the bulge angle found for the parametric S-model which we used as a prior for our non-parametric fit. Our parametric prior fit produces a bulge angle of $19.8^\circ$, comparable to the $19.6^\circ$ angle in S17  which uses a similarly dispersed RC luminosity function. Some older works, such as WG13,  used a much broader luminosity function. The dependence of the major axis angle on the broadness of the luminosity function has been commented on by  S17 for the VVV catalogue as well as by 
\cite{StanekModellingGalacticBar1997}
 for the RC in the OGLE data. The narrower luminosity function which we have used is  more consistent with recent measured intrinsic RC magnitude dispersions \citep{HallTestingasteroseismologyGaia2019,ChanemphGaiaDR2parallax2019}.}

The X-arms are visible at $|z| > 0.319$~kpc.
Although WG13 had a similar result for $z<0$,
they had significant gaps in their reconstruction for
$z>0.263$~kpc. However, they filled in these gaps by assuming eight-fold symmetry.
As we did not have this problem, we did not need to make any symmetry assumptions.
Our less restrictive symmetry assumptions have also allowed us to uncover the presence 
of a 
 spiral arm structure in front of the bulge, which is visible in the deconvolved density (left panels of Fig.~\ref{fig:spiralarmdensity}) at $|z| < 500$~pc at $x\sim-3$~kpc. 
 We also found a 
  spiral arm structure behind the bulge which is  visible at all $|z| < 1$~kpc at $x\sim3$~kpc. 
 
\begin{figure*}
\centering
    \includegraphics[width=\textwidth]{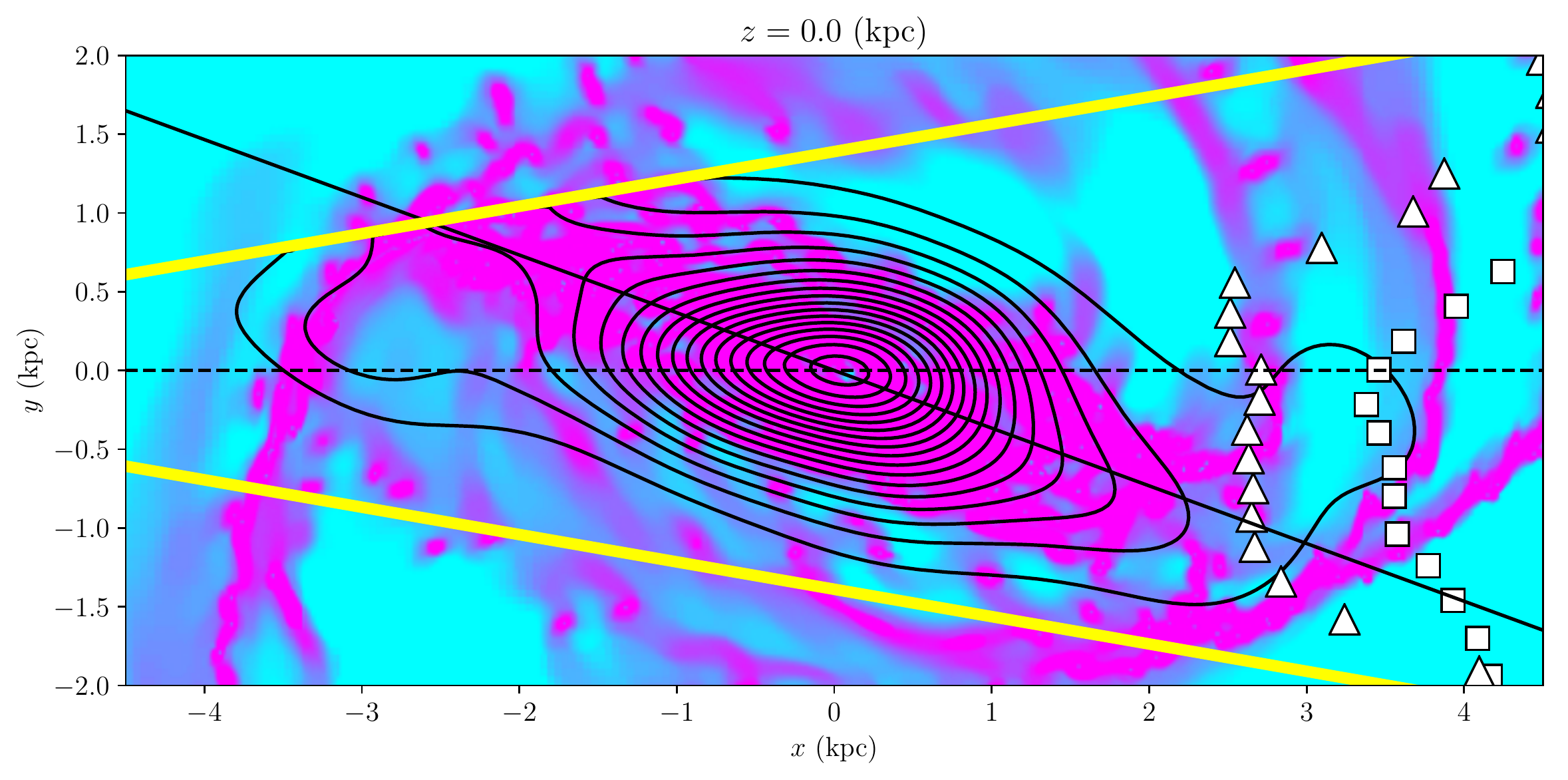}
    \caption{VVV deconvolved stellar density (black contours) as compared to the simulated inner galaxy gas distribution of \protect\cite{Renaud2013}. The Sun is located at $(x,y,z)=(-8.0,0.0,0.0)$. The location of the spiral arm structure behind the bulge falls between 
    the simulation (white squares)
    and VVV data analysis predictions  (white triangles)
    of \protect\cite{Gonzalez2018StructureBehindBar}.  The yellow lines show $l = \pm 10^{\circ}$. The spiral arm structure at the end of the bulges are offset from the bulge major axis (black solid line).) 
    \label{fig:spiralarmcomparison}
    }
\end{figure*}

The positions of the spiral arm structures in Fig. \ref{fig:spiralarmcomparison} are consistent with the inner galaxy of the simulated gas distribution of \cite{Renaud2013}\footnote{\url{http://www.astro.lu.se/~florent/mw_large.php}}. The location of the spiral arm structure from \cite{Gonzalez2018StructureBehindBar} (white triangles in Fig.~\ref{fig:spiralarmcomparison}) are closer to the Sun than predicted by our model. This is likely because we have only considered fields which are not heavily effected by extinction and crowding. Additionally, the $K_s$ magnitudes of the VVV stars in \cite{Gonzalez2018StructureBehindBar} have the photometric zero-point calibrated to the CASU aperture photometry catalogues, which is not consistent with the corrected zero-point magnitudes we have used. 

In order to evaluate the apparent magnitude of the spiral arm structure  in front of the bulge, we note that as illustrated in Fig.~\ref{fig:luminosityfunction}, 
$M_{K_{s}} \approx -1.53$ mag for the RC. Also, as can be seen from Fig.~\ref{fig:spiralarmcomparison}, the distance of the spiral arm structure in front of the bulge from the Sun is 5~kpc. We can then use the standard relation
\begin{equation}
    K_s-M_{K_s}=5\log_{10}s+10
\end{equation}
where $s$ is in kpc. Substituting in the above values and solving  gives $K_{s}\approx 11.96$ mag. As \cite{Gonzalez2018StructureBehindBar} only used data with $K_s\gtrsim 12$~mag, they would not have been sensitive to the spiral arm structure in front of the bulge.

As can be seen in Fig.~\ref{fig:spiralarmcomparison}, the spiral arm structure at positive $x$ connects to the bulge below the major axis, while the spiral arm structure at negative $x$ connects to the bulge  above the major axis. 
As can be seen by comparing, for example, the $z=0$~kpc left and $z=-0.8361$~kpc right hand side panels of  Fig~\ref{fig:spiralarmdensity}, the spiral arm structures are offset from the bulge major axis on the same side as the high $|z|$ X-arm maxima.

\begin{figure*}
\centering
 \includegraphics[width=\textwidth]{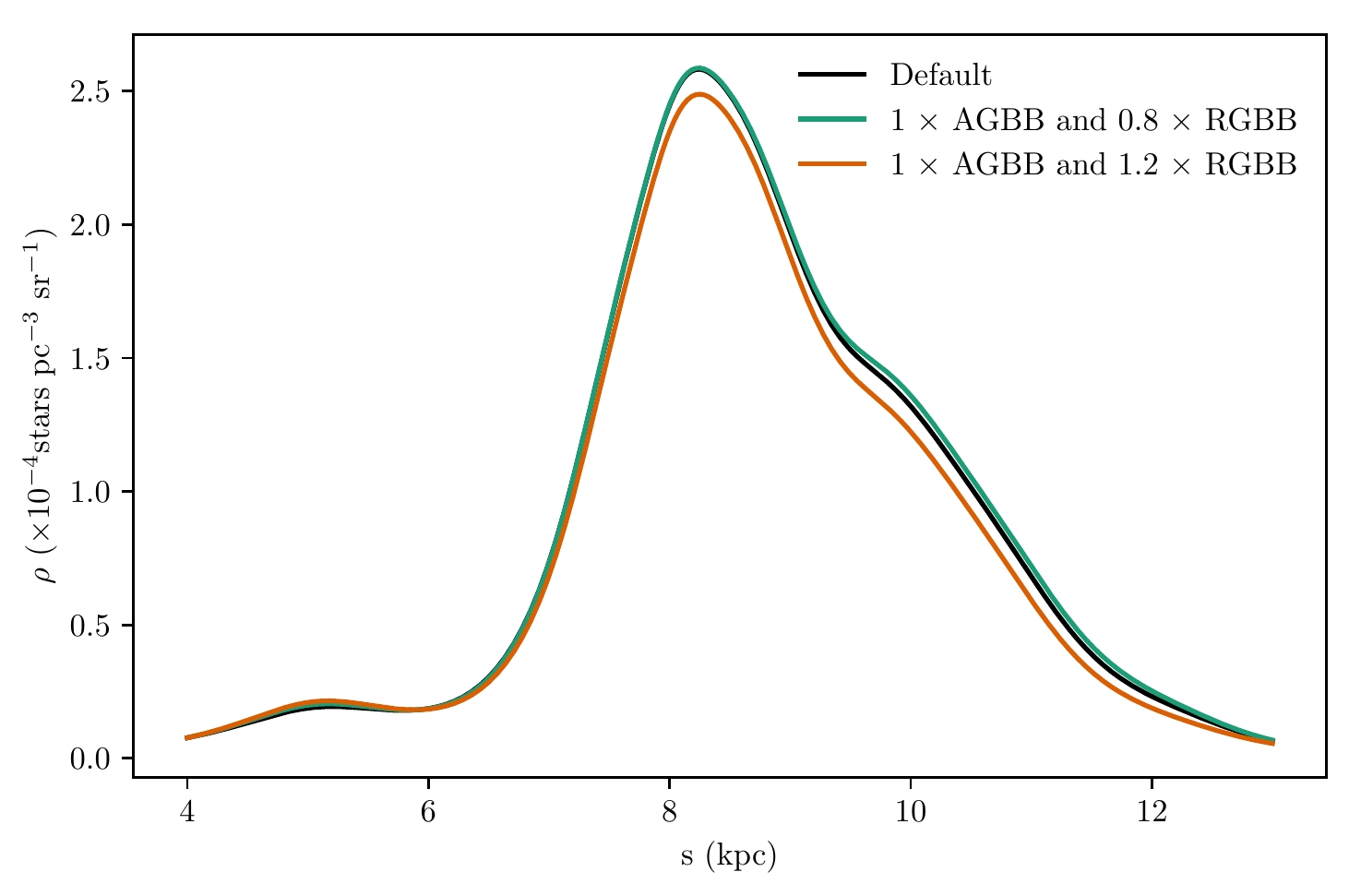}
   \includegraphics[width=\textwidth]{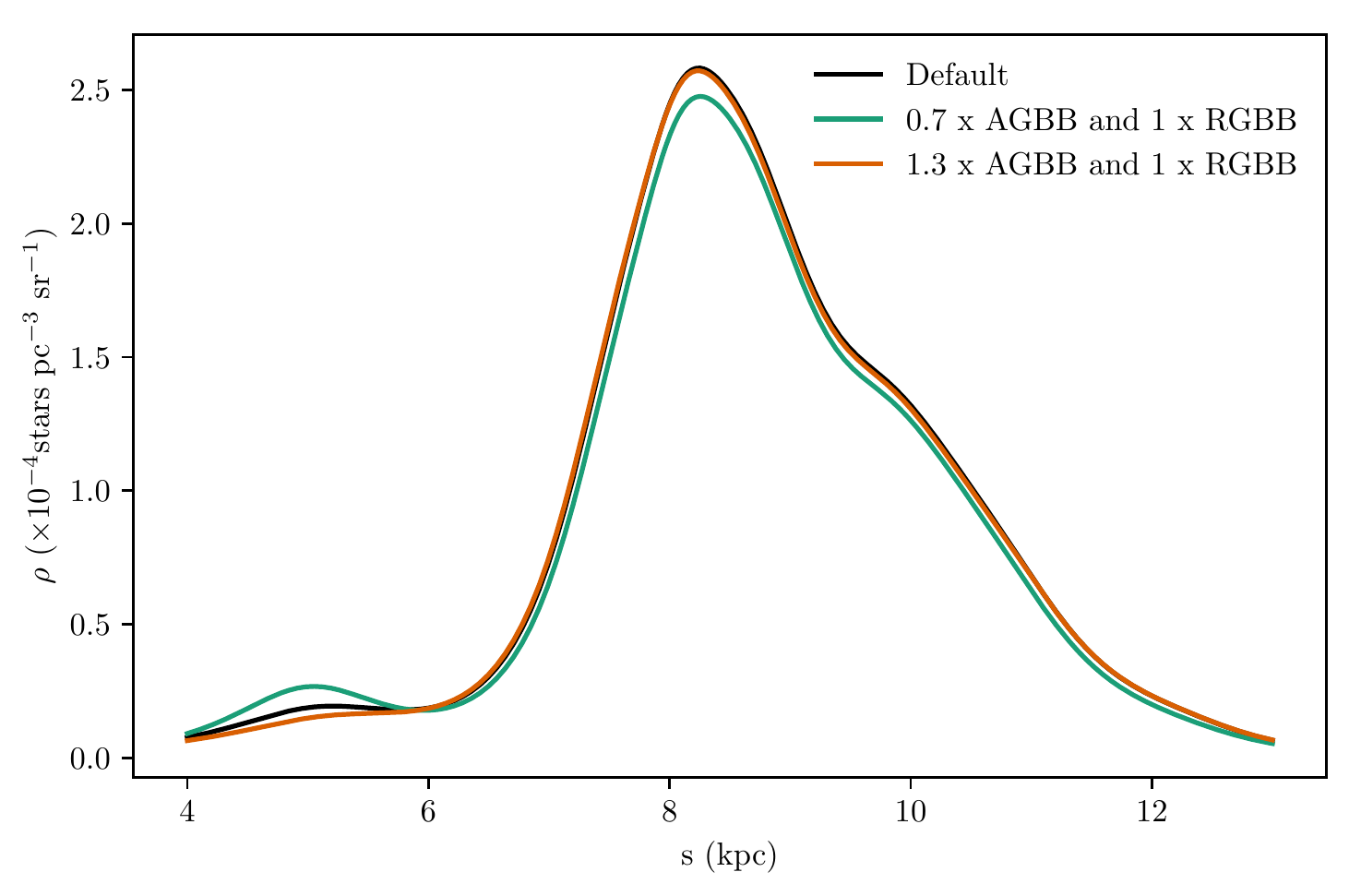}
    \caption{\newnewtext{Line of sight plots for $(l,b)=(-8.1^\circ,-2.9^\circ)$ of the reconstructed density for different luminosity functions. Top panel: 
    The `Default'' case is the reconstructed density from the VVV data,  when our luminosity function shown in Fig.~\ref{fig:luminosityfunction} is used. The other cases correspond to the default luminosity function but with either the RGBB component increased by multiplying by a factor of 1.2 or decreased by multiplying by a factor of 0.8. 
    Bottom panel: Similar to the top panel except the AGBB component has been increased by multiplying by a factor of 1.3 and decreased by multiplying by a factor of 0.7.
    }
    \label{fig:systematics}
    }
\end{figure*}

 As noted by \cite{Gonzalez2018StructureBehindBar}, the red giant branch bump (RGBB) of the bulge has a similar $K_s$ to the feature behind the bulge. A mismodelling of the RGBB might explain some of the signal at $z$ far from the Galactic Midplane seen in the right hand side panels of Fig.~\ref{fig:spiralarmdensity}, where we may not expect a spiral-arm-like density.
 We examined the effects of changing the asymptotic giant branch bump (AGBB) and RGBB contribution in Fig.~\ref{fig:systematics}. 
 \newnewtext{We used the line of sight along  $(l,b)=(-8.1^\circ,-2.9^\circ)$ as in that direction the density of the background (and foreground) structures becomes significant with respect to the bulge.
 This can be seen for example from Fig.~\ref{fig:spiralarmcomparison} and also 
  Fig.~3 of \citet{Gonzalez2018StructureBehindBar}. 
  Based on the results of \citet{Nataf2011} we used a 20\% variation  in the amplitude of the RGBB luminosity peak and a 30\% variation in the amplitude of the AGBB luminosity peak.
   Note that RC stars that are in the behind the bulge feature have a similar apparent magnitude to the RGBB stars that are in the main population of the bulge. While RC stars in the feature in front of the bulge have a similar apparent magnitude to the AGBB stars in the main population of the bulge. Therefore, increasing/decreasing the density of the  feature behind the bulge can account for an underestimation/overestimation of the RGBB in the luminosity function. Increasing/decreasing the density of the feature in front of the bulge can account for an underestimation/overestimation of the AGBB in a similar way.
  As can be seen in Fig.~\ref{fig:systematics}, in the default case there is a maximum at around $s=5$~kpc which we interpret as the feature in front of the bulge. We can reduce the magnitude of that maximum by increasing the size of the AGBB peak by 30\% as seen with the orange line in the bottom panel. But as the feature is still there we can conclude the detection of the feature in front of the bulge is robust to the uncertainty of the amplitude of the AGBB luminosity function peak. While at $s=10$~kpc there is a saddle point in the default case  which corresponds to the feature behind the bulge. This saddle point is still clearly there in the case where we increase the RGBB size by 20\% in the top panel. So this shows that the feature behind the bulge is robust to the uncertainty of the height of the RGBB luminosity function peak. 
  }

\cite{PaperII} were only concerned with the morphology of the bulge and so they removed the apparent features behind and in front of the bulge by employing a non-parametric estimate of the background. Although, there where still some remnants of the feature behind the bulge which they then manually removed.  

\begin{figure*}
\centering
    \includegraphics[width=0.9\textwidth]{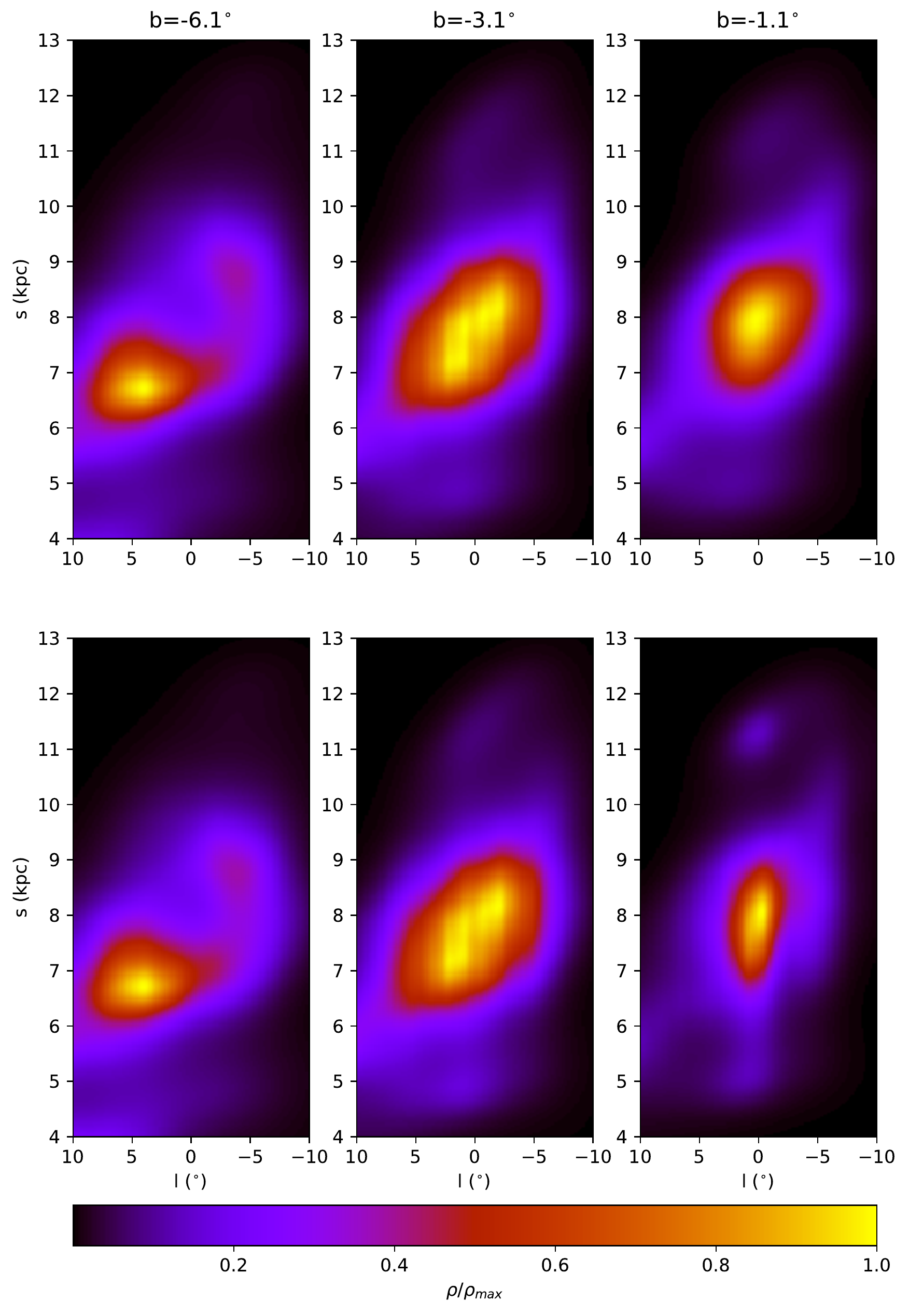}
    \caption{\newtext{Density reconstruction plots. The top three panels are for our standard mask shown in the right hand panel of  Fig.~\ref{fig:vvvmasks} and the bottom three panels are for an alternative which masked out $\left|b\right|<1^\circ$.
    }
    \label{fig:masksystematics}
    }
\end{figure*}

\newtext{We also examined the impact of the mask choice. We did this by redoing our analysis with only the $\left|b\right|<1^\circ$ region masked. As can be seen from the bottom right hand panel of Fig.~\ref{fig:masksystematics}, the reconstruction with this less conservative mask has a prominent finger to  sun feature close to the galactic plane. As can be seen from Fig.~\ref{fig:vvvmasks}, this is likely to be an artefact as this is a region with high photometric error  and this region also had the greatest zero point offset when matching to 2MASS.}

\section{Conclusions}\label{sec:Conclusion}
We have used a non-parametric method incorporating maximum entropy and smoothness regularisation to deconvolve the density distribution of bulge stars in the VVV MW-BULGE-PSFPHOT catalogue. We have observed a new morphological feature in the inner region of the Milky Way: a feature  $\sim3$~kpc in front of the Galactic centre consistent with a foreground spiral arm. 
We also confirmed
the observation made by \cite{Gonzalez2018StructureBehindBar}
of a spiral arm structure $\sim3$~kpc behind the Galactic centre.
The spiral arm structures are connected on opposite sides of the major axis. \newnewtext{We showed that detection of the feature in front of the bulge and the detection of the feature behind the bulge are both robust to the uncertainties of the AGBB and RGBB luminosity peak sizes.}\\

\noindent {\bf Data Availability: } The data underlying this article will be shared on reasonable request to the corresponding author.

\section*{Acknowledgements}
We thank Iulia Simion, Victor Debattista, and an anonymous referee for helpful discussions and comments. Also, we are  grateful to Elena Valenti for giving us early access to the MW-BULGE-PSFPHOT VVV catalogue.
This work was made possible by the use of the Research Compute Cluster (RCC) facilities at the University of Canterbury.
The following software packages were used: \textsc{astropy}, \textsc{matplotlib}, \textsc{numpy}, \textsc{pylbfgs}, and \textsc{scipy}.

\bibliographystyle{mnras}
\bibliography{references} %

\appendix

\section{Analytic Likelihood Gradient}
\label{Appendix:grad}
The analytic gradient of $\ln \mathcal{L}$ were determined as follows. We take $\rho$ to be a field over $(s,l,b)$ so that a single density value, $\rho_\delta = \rho(s,l,b) = \rho_{h,j,k}$ where $\{h,j,k\}\in \{s,l,b\}$. Each of the $\rho_\delta$ represents a free parameter in our model. The gradient of $\ln \mathcal{L}$ with respect to $\rho_\delta$ is then
\begin{equation}
    \frac{\partial }{\partial \rho_\delta} \ln \mathcal{L} = \frac{\partial }{ \partial  \rho_\delta} \ln\mathcal{L}_{\rm{P}} + \frac{ \partial  }{ \partial \rho_\delta} \ln\mathcal{L}_{\rm MEM} + \frac{ \partial }{ \partial \rho_\delta} \ln\mathcal{L}_{\rm{smooth}}
\end{equation}
where $\ln\mathcal{L}_{\rm{P}}$ is the Poisson log-likelihood, $\ln\mathcal{L}_{\rm MEM}$ is the maximum entropy penalty term and $\ln\mathcal{L}_{\rm{smooth}}$ is the smoothness penalty term. The gradient of the Poisson term is then 
\begin{equation}
    \frac{\partial}{\partial \rho_\delta}\ln\mathcal{L}_{\rm{P}} = \sum_{\{i,j,k\}\in \{K_s,l,b\}} \left( \frac{n_{i,j,k}}{N_{i,j,k}} - 1 \right)\left(\frac{\partial N}{\partial \rho_\delta}\right)_{i,j,k}
    \label{eq:poissonderivative}
\end{equation}
where the derivative of the predicted counts, $N$, is determined by differentiating Eq.~\eqref{eq:stellarstatistics} with respect to $\rho_\delta$ so that
\begin{multline}
    \left(\frac{\partial N}{\partial \rho_\delta}\right)_{i,j,k} = \\
    \Delta \Omega \Delta K_s \left(\int^{13}_{4} \frac{\partial\rho}{\partial \rho_\delta} \Phi\left(K_{s} - 5\log s - 10\right) s^2 \, {\rm d}s\right)_{i,j,k}.
    \label{eq:modelderivative}
\end{multline}
The integral above can be approximated by the midpoint rule:
\begin{multline}
    \left(\frac{\partial N}{\partial \rho_\delta}\right)_{i,j,k} = \\
    \Delta \Omega \Delta K_s \Delta s\left(\sum_{h\in s} \left(\frac{\partial\rho}{\partial \rho_\delta}\right)_{h,j,k} \Phi_{h,i,j,k} s_h^2 \right)_{i,j,k}
    \label{eq:modelderivative1}
\end{multline}
Where $\Phi_{h,i,j,k}$ is the discretised version of the luminosity function which needs an index for
$s$, $K_s$, $l, $ and b.
Since $\rho_\delta$ is single value at $(h,j,k) =( h^\prime, j^\prime,k^\prime)$ in the field $\rho$, then $\frac{\partial \rho}{\partial \rho_\delta} = \delta_{hh^\prime} \delta_{jj^\prime} \delta_{kk^\prime}$, where the $\delta$ here are the Kronecker delta. Substituting this into  Eq.~\eqref{eq:modelderivative1} and simplifying gives 
\begin{equation}
    \left(\frac{\partial N}{\partial \rho_\delta}\right)_{i,j,k} = \Delta \Omega \Delta K_s \Delta s \,    \delta_{jj^\prime} \delta_{kk^\prime} \Phi_{h^\prime,i,j,k} s_{h^\prime}^2 \,.
    \label{eq:modelderivativesimplified}
\end{equation}
Substituting Eq.~\eqref{eq:modelderivativesimplified} into Eq.~\eqref{eq:poissonderivative} gives the final form of the Poisson component of the gradient as
\begin{equation}
    \frac{\partial}{\partial \rho_\delta}\ln\mathcal{L}_{\rm{P}} = \Delta \Omega \Delta K_s \Delta s\sum_{i\in K_s} \left( \frac{n_{i,j^\prime,k^\prime}}{N_{i,j^\prime,k^\prime}} - 1 \right)   \Phi_{h^\prime,i,j^\prime,k^\prime} s_{h^\prime}^2.
\end{equation}
The gradient for the Poisson component is equal to zero when $n = N$, so that the log-likelihood has an extremum when the model, $N$, and the data, $n$, are equal.

The gradient of the maximum entropy penalty was determined by taking the derivative of Eq.~\eqref{eq:maxentdef} with respect to $\rho_\delta$ and applying the chain rule

\begin{multline}
    \frac{\partial}{\partial \rho_\delta}\ln\mathcal{L}_{\rm MEM} = \\\frac{\partial}{\partial \kappa}\left(-\lambda \sum_{\{h,j,k\} \in \{s,l,b\}} \left( 1 - \kappa_{h,j,k} + \kappa_{h,j,k} \ln \kappa_{h,j,k} \right)\right) \frac{\partial \kappa}{\partial \rho_\delta}
    \label{eq:MEMgrad}
\end{multline}
where $\kappa= \frac{\rho}{\rho_{\rm prior}}$ is the ratio between the density field and a smooth prior estimate, $\rho_{\rm prior}$. Evaluating the derivative in Eq.~\eqref{eq:MEMgrad} and substituting in $\frac{\partial \kappa}{\partial \rho_\delta} = \frac{1}{(\rho_{\rm prior})_{\delta}}$ gives
\begin{equation}
    \frac{\partial}{\partial \rho_\delta}\ln\mathcal{L}_{\rm MEM} = - \frac{ \lambda}{(\rho_{\rm prior})_{\delta}} \ln \left \lbrack \frac{\rho_\delta}{(\rho_{\rm prior})_{\delta}} \right \rbrack,
    \label{eq:MEMgradsimplified}
\end{equation}
which equals zero (thus giving the extremum) when $\rho_\delta = (\rho_{\rm prior})_\delta$. It follows from this equation that 
\begin{equation}
\left.    \frac{\partial^2}{\partial \kappa_\delta^2}\ln\mathcal{L}_{\rm MEM}\right|_{\kappa_\delta
=1}=-\lambda.
    \label{eq:MEMgradgradsimplified}
\end{equation}
We can then evaluate the expected deviation from the prior using the standard Gaussian approximation for estimating errors in  maximum likelihood:
\begin{equation}
\sigma_{\rm MEM}\equiv\sigma_\kappa=\left(- \left.    \frac{\partial^2}{\partial \rho_\delta^2}\ln\mathcal{L}_{\rm MEM}\right|_{\rho_\delta = (\rho_{\rm prior})_\delta}\right)^{-1/2} ={1\over \sqrt{\lambda}}
    \label{eq:MEMsigma}
\end{equation}

The gradient of the smoothing term was obtained by direct differentiation of the last three lines of Eq.~\eqref{eq:likelihoodmaxent} so that

\begin{equation}
\label{eq:PartialDerivativeofCurvature}
\begin{split}
\frac{ \partial }{ \partial \rho_\delta} & \ln\mathcal{L}_{\rm{smooth}} = \\
&-\frac{\eta_s}{\rho_\delta}\left\lbrack \left( \ln \rho_{h^{\prime}-2,j^{\prime},k^{\prime}} - 4\ln  \rho_{h^{\prime}-1,j^{\prime},k^{\prime}} + 6\ln \rho_{h^{\prime},j^{\prime},k^{\prime}} \right. \right. \\
&\left. \left.- 4\ln \rho_{h^{\prime}+1,j^{\prime},k^{\prime}} + \ln \rho_{h^{\prime}+2,j^{\prime},k^{\prime}} \right) \right\rbrack\\-
&\frac{\eta_l}{\rho_\delta}\left\lbrack \left( \ln \rho_{h^{\prime},j^{\prime}-2,k^{\prime}} - 4\ln  \rho_{h^{\prime},j^{\prime}-1,k^{\prime}} + 6\ln \rho_{h^{\prime},j^{\prime},k^{\prime}} \right. \right. \\
&\left. \left.- 4\ln \rho_{h^{\prime},j^{\prime}+1,k^{\prime}} + \ln \rho_{h^{\prime},j^{\prime}+2,k^{\prime}} \right) \right\rbrack\\-
&\frac{\eta_b}{\rho_\delta}\left\lbrack \left( \ln \rho_{h^{\prime},j^{\prime},k^{\prime}-2} - 4\ln  \rho_{h^{\prime},j^{\prime},k^{\prime}-1} + 6\ln \rho_{h^{\prime},j^{\prime},k^{\prime}} \right. \right. \\
&\left. \left.- 4\ln \rho_{h^{\prime},j^{\prime},k^{\prime}+1} + \ln \rho_{h^{\prime},j^{\prime},k^{\prime}+2} \right) \right\rbrack
\end{split}
\end{equation}
It is easy to check by substitution that this equation is equal to zero when $ \rho_{h,j,k} $ is an exponential function of the form:
\begin{equation}
\label{eq:ExpFunction}
\rho_{h,j,k}=A \exp(A_h h+A_j j+A_k k )
\end{equation}
where $A$, $A_h$, $A_j$, and $A_k$ are constants. Similarly  to the MEM case, it follows from differentiating Eq.~\eqref{eq:PartialDerivativeofCurvature} and evaluating the result using \eqref{eq:ExpFunction} that the relative deviation from an exponential function is given by: 
\begin{equation}
    \sigma_{\rm curvature}\equiv {\sigma_{ \rho}\over \rho_\delta}={1\over\sqrt{6\eta}}.
    \label{eq:SigmaCurvature}
\end{equation}

\newtext{
\section{Simulated Data}
\label{Appendix:sim}
We constructed a simulated Milky Way population from the thin disc and thick disc described in Sec.~\ref{sec:Method}. We also included our non-parametric estimate of the bulge density.
  To generate the simulated population, we used
\begin{multline}
    \label{eq:stellarstatisticssim}
    N\left(K_{s}, l, b\right) = \Delta \Omega\Delta K_s\\ \times \sum_{i}\int^{\infty}_{0} \rho_{i}\left(s, l, b\right) \Phi_{i}\left(K_{s} - 5\log s - 10\right) s^2  ds
\end{multline}
where $\rho$ is the density and $\Phi$ is the luminosity function and the sum is over the
thick disk, the thin disk, and the bulge. This predicts the combined star counts in each $(K_s,l,b)$ voxel. We then simulated a population of stars by drawing a Poisson random value from the binned simulation model. }

\newtext{The normalisations we used for each of the three components were multiplied by the same constant chosen so that  the total number of stars in the unmasked region and in $12 < K_s < 14$ matches the number of stars in the VVV PSF catalogue. The luminosity function we used for the bulge in the simulation is the same as the one we used in our fitting procedure to the VVV data and is plotted in Fig.~\ref{fig:luminosityfunction}.
 }

\bsp	%
\label{lastpage}
\end{document}